\documentclass{aa}
\usepackage{times}
\usepackage{amssymb}
\usepackage{epsfig}
\usepackage{array}
\usepackage{amsmath}

\newcommand{\Lya}{Ly$\alpha$}

\title{The extended Lyman-$\alpha$ emission surrounding the $z=3.04$
  radio-quiet QSO1205-30: Primordial infalling gas illuminated by the
  quasar?\thanks{Based on observations made with ESO Telescopes at the
  Paranal Observatory under programme ID 64.O-0187} }

\author{M.~Weidinger\inst{1,2}
\and P.~M\o ller\inst{2}
\and J.~P.~U.~Fynbo\inst{1,3}
\and B.~Thomsen\inst{1}
}

\institute{ 
Institute of Physics and Astronomy, University of Aarhus, Ny
Munkegade, DK-8000 {\AA}rhus C, Denmark
\and
European Southern Observatory, Karl-Schwarzschild-Stra{\ss}e 2,
D-85748 Garching bei M\"unchen, Germany
\and
Astronomical Observatory, University of Copenhagen, Juliane Maries Vej
30, DK-2100 Copenhagen {\O}, Denmark
} 
\mail{michaelw@phys.au.dk}
\offprints{M.~Weidinger}
\date{Received / Accepted }

\titlerunning{The extended Lyman-$\alpha$ emission around the
  radio-quiet QSO1205-30}
\authorrunning{M.~Weidinger et al.}

\begin{document} 
\abstract{We present spectroscopic observations obtained with the FORS1
instrument on the ESO VLT under good seeing conditions of the
radio-quiet $z_{\rm em}=3.04$ quasar \object{Q1205-30} and its associated
extended {\Lya} emission. The extended {\Lya} emission was originally
found in a deep narrow band image targeting a $z_{\rm abs} \approx
z_{\rm em}$ Lyman-limit system in the spectrum of the QSO. Using
spectral point-spread function fitting to subtract the QSO spectrum, we
clearly detect the extended {\Lya} emission as well as two foreground
galaxies at small impact parameters ($2.12\pm 0.04$ and $2.77\pm 0.07$
arcsec). The redshifts of the two foreground galaxies are found to be
$z=0.4732$ and $z=0.865$. We determine the redshift and velocity
profile for the extended {\Lya} emission, and analyzing the velocity
offsets between eight QSO emission lines we refine the quasar redshift
determination. We use the new redshifts to infer the geometry of the
complex. We find that the extended {\Lya} emission is clearly
associated with the quasar. A {\Lya} luminosity of $5.6\times 10^{43}
\text{~erg~s}^{-1}$ places this extended emission at the high luminosity
end of the few previous detections around radio-quiet quasars. The
extended {\Lya} emission is best explained by hydrogen falling into the
dark matter halo inhabited by the quasar.

 }

\maketitle

\section{Introduction}

Quasar host galaxies are visible tracers of the close environment of
this powerful type of active galactic nuclei (AGN). The feeding of the
central engine from the host galaxy, and the feedback of the quasar to
the host are important unknown factors in current numerical models,
which need to be understood. The study of quasar host galaxies is
difficult, because of the high contrast between the bright point-source
quasar and the faint, extended host galaxy. Surveys of AGN host
galaxies have primarily targeted radio-loud quasars (RLQs) and radio
galaxies (RGs) (e.g.~Lehnert et al.~1992, 1999; Reuland et al.~2003;
S\'anchez \& Gonz\'alez-Serrano 2003), despite the fact that the
majority of quasars are radio-weak or radio-quiet. In such surveys it
has been found that RLQs and RGs at low $z$ (here taken to mean $z<1$)
reside in luminous elliptical galaxies, while radio-quiet quasars
(RQQs) are found in both elliptical and early spiral galaxies. At
intermediate redshifts ($1\lesssim z \lesssim 2$) galaxies in general
appear to have more disturbed morphologies, making it difficult to
apply the simple classification of ``ellipticals'' and
``spirals''. Studies have found that host galaxies of RQQs are
$1-2$~mag fainter than hosts of RLQs with similar luminosity (Falomo et
al.~2001; Kukula et al.~2001). Similarly, in a study of quasars
out to $z=2.1$ the black holes of RLQs are typically found to be $45\%$
more massive than their radio-quiet counterparts (McLure \& Jarvis
2004). At $z>2$ the cosmological surface brightness dimming makes it
increasingly difficult to make secure detections of the host
galaxies. For RLQs some $20$ examples of $z>2$ hosts are seen (Hu et
al.~1991; Heckman et al.~1991a,b; Steidel et al.~1991; Lehnert et
al.~1992, 1999; Wilman et al.~2000), but only few surveys have targeted
RQQs and with limited success (Bremer et al.~1992; Lowenthal et
al.~1995; Fynbo et al.~2000a; M{\o}ller et al.~2000; Ridgway et
al.~2001; Bunker et al.~2003). The relatively faint hosts of RQQs
compared to RLQs makes them more difficult to detect. A promising
method is narrow-band imaging tuned to the {\Lya} line at the quasar
redshift (Hu {\&} Cowie 1987; Hu et al.~1996). Haiman {\&} Rees (2001)
predict that gas enshrouding a quasar between redshifts 3 and 8 would
be photoionized by the quasar UV emission and should be detectable at a
surface brightness of $10^{-18}$ to $10^{-17} \text{~erg~s}^{-1}
\text{~cm}^{-2} \text{~arcsec}^{-2}$ in the {\Lya} line. These limits
have only been reached for very few surveys.

In this paper we report on a spectroscopic study of the sightline
towards the radio-quiet quasar \object{Q1205-30} at $z=3.04$ and its associated
extended {\Lya} emission detected by Fynbo et al.~(2000b) (hereafter
Paper I).

The paper is organized as follows: In Sect.~\ref{sec:q1205:2} we
present the observations and data reductions. We continue in
Sect.~\ref{sec:q1205:3} with the results of our analysis. In
Sect.~\ref{sec:abslines} we discuss the foreground galaxies, and we
present a simple model for the extended {\Lya} emission in
Sect.~\ref{sec:model}. We end in Sect.~\ref{sec:q1205:5} with a
discussion of the origin of the extended emission. Unless stated
otherwise we will use $\Omega_{\rm m}=0.3$, $\Omega_\Lambda = 0.7$,
$H_0 = 100 h \text{ km s}^{-1}\text{Mpc}^{-1} = 70\text{ km
s}^{-1}\text{Mpc}^{-1}$. In this model a redshift of $3.04$ corresponds
to a luminosity distance $D_L = 25.8$~Gpc and a distance modulus of
$47.1$. One arcsec on the sky corresponds to a projected distance of
$7.67$ proper kpc and the lookback time is $11.4$ Gyr ($84.5\%$ of the
time since Big Bang).

\section{Observations and data reduction}\label{sec:q1205:2}

The observations were carried out with the Unit Telescope 1 (Antu) of
the ESO Very Large Telescope (VLT) on March 4--5, 2000, under
photometric and good seeing conditions. The data were acquired with the
FOcal Reducer/low dispersion Spectrograph (FORS1) instrument in Multi
Object Spectroscopy (MOS) mode with the red G600R and blue G600B grisms
as a part of a larger campaign (see Fynbo et al.~2001). The slitlet
used was $1.2$~arcsec wide and $20$~arcsec long, and the positions on
the CCD resulted in a wavelength coverage of approximately
$4000-6000$~{\AA} (G600B) and $5800-7800$~{\AA} (G600R). The slitlets
had position angles of $67.20^\circ$ (PA1) and $7.90^\circ$ (PA2) East
of North centred on the quasar (see below). For the observations we
used the standard resolution collimator. During observations with the
G600B grism the CCD was binned $2\times 2$, and resulting in a pixel
size of $0.4$~arcsec by $2.4$~{\AA} for G600B and $0.2$~arcsec by
$1.2$~{\AA} for G600R. The seeing in the combined science frames was
measured to be $0.9$~arcsec (PA1) and $0.7$~arcsec (PA2) at a
wavelength of $\sim 5100$~{\AA}, leading to spectral resolutions of
$5$~{\AA} (PA1) and $4$~{\AA} (PA2). The exposure times are given in
Table~\ref{tab:exp}.

\begin{table}[tbp]
\caption{{Observation log.}}\label{tab:exp}
\centering
\begin{tabular}{c c c c c}
\hline\hline
    & PA     & Grism & Seeing & Exp-time \\\hline
PA1 & $67.20^\circ$ & G600B & 0\farcs91 & $4\times 1800$ s \\
PA1 & $67.20^\circ$ & G600R & 0\farcs69 & $2\times 1800$ s \\
PA2 &  $7.90^\circ$ & G600B & 0\farcs69 & $4\times 1800$ s \\
PA2 &  $7.90^\circ$ & G600R & 0\farcs73 & $3\times 1800$ s \\\hline
\end{tabular}
\end{table}

\begin{figure}[tbp]
\centering\epsfig{file=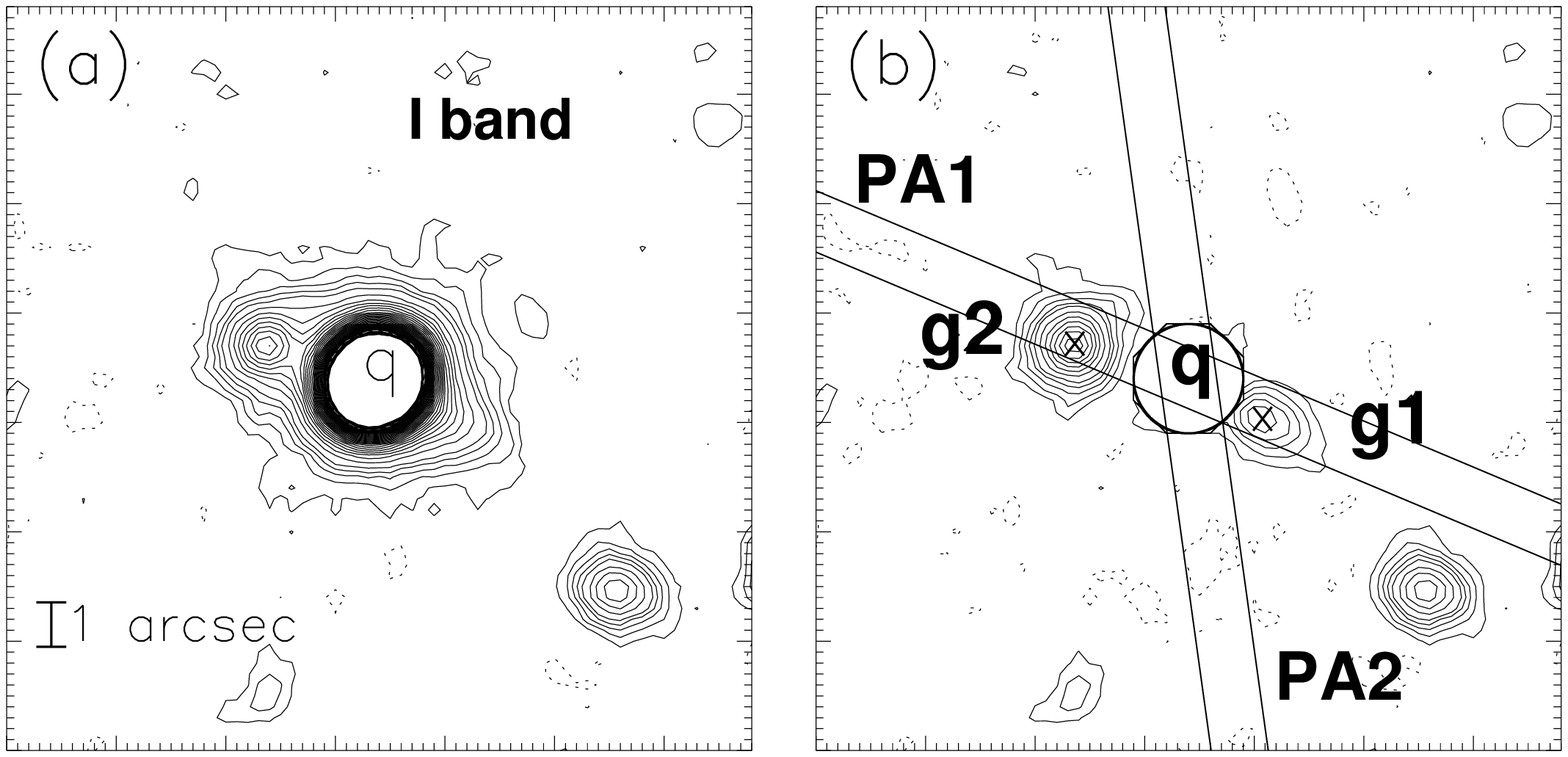,width=8.5cm}
\centering\epsfig{file=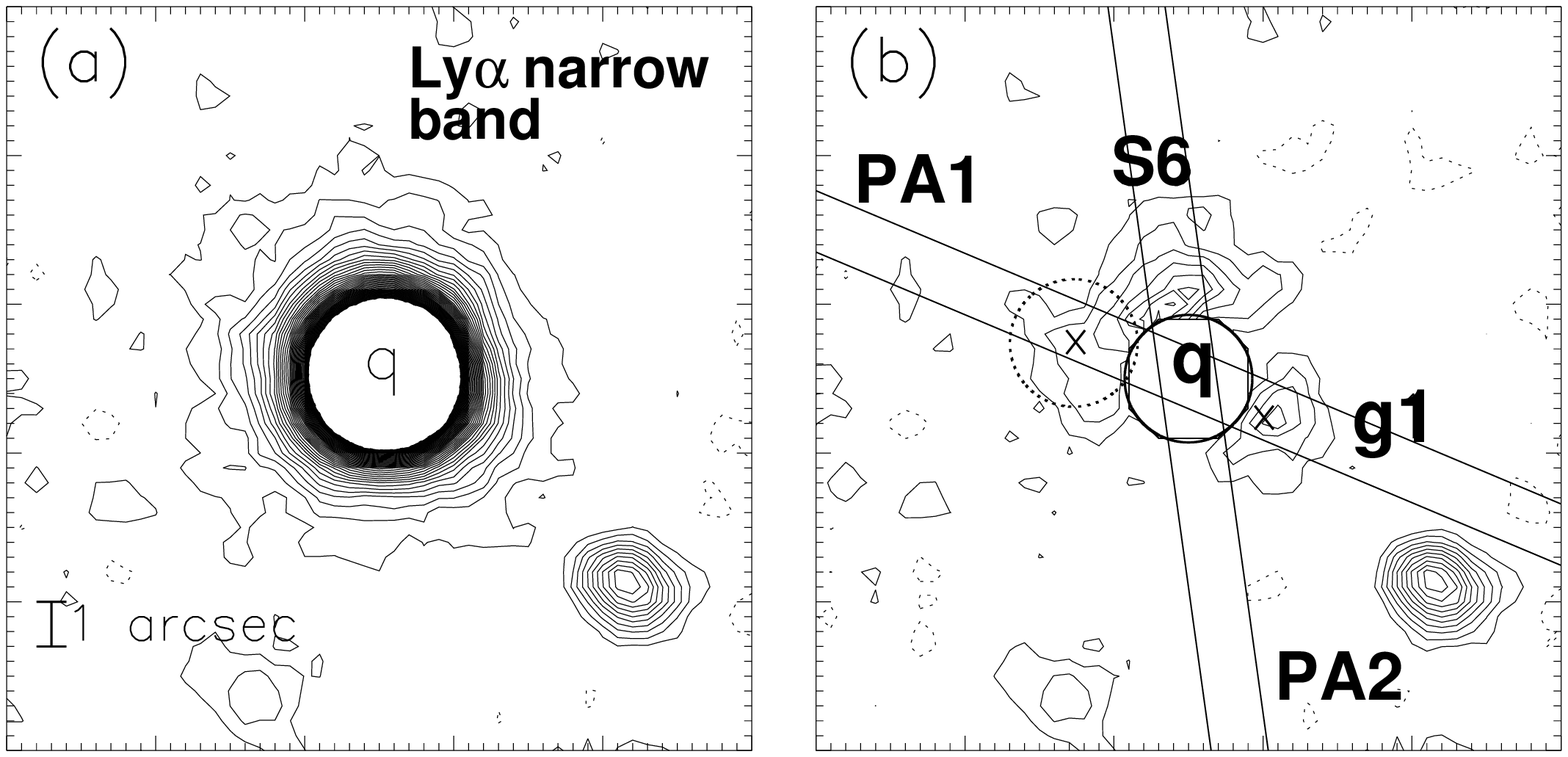,width=8.5cm}
\caption{ The $18\times 18$~arcsec$^2$ field of \object{Q1205-30}. {\it Upper
panels:} Contour plot of the $I$-band image before (a) and after (b)
PSF subtraction of the quasar. North is up and East is left. {\it
Lower panels:} Contour plot of the narrow band image before (a) and
after (b) PSF subtraction of the quasar. North is up and East is
left. Taken from Paper I. }\label{fig:field}
\end{figure}

Contour plots of the $18\times 18$~arcsec$^2$ field of \object{Q1205-30} imaged
in $I$ and {\Lya} narrow band is shown in Fig.~\ref{fig:field} taken
from Paper I. We shall here follow the naming convention of that paper,
i.e.~g1 is the blue galaxy SW of the quasar, g2 is the red galaxy NE of
the quasar, and S6 is the extended {\Lya} emission N and NE of the
quasar. The projected distances from the QSO are $2.12\pm 0.04$~arcsec
for g1 and $2.77\pm 0.07$~arcsec for g2. For PA1 the slitlet covers g1,
g2, the QSO and a part of S6, whereas for PA2 the slitlet covers the
central part of S6 and the QSO.

\subsection{Basic reductions}
The individual science frames were bias subtracted using standard
techniques. The flat fielding was done by first filtering the flat
fields along the dispersion axis with a $61\times 1$~pixels
($146$~{\AA} long) median filter for the G600B grism, and a $121\times
1$~pixels ($145$~{\AA} long) median filter for the G600R grism. Then
the flat fields were normalized by dividing the unfiltered flat fields
by the filtered ones, and finally we divided the science frames by
these normalized flat fields. In order to obtain a mean sky spectrum
spatial bins on both sides of the quasar spectrum were filtered using a
$1\times 13$~pixels median filter (i.e.~only filtering along the
spatial axis) to remove cosmic ray hits and averaged. Regions used for
determining the sky spectrum were never closer than $5$~arcsec to the
QSO on the side of the extended emission and $4$~arcsec on the opposite
side. The mean sky spectrum was expanded to a two dimensional spectrum
by duplicating the 1D spectrum and subtracted from the unfiltered
science frame.

\subsection{Spectral extractions}\label{sec:extraction}
\begin{figure*}[tbp]
\centering
\epsfig{file=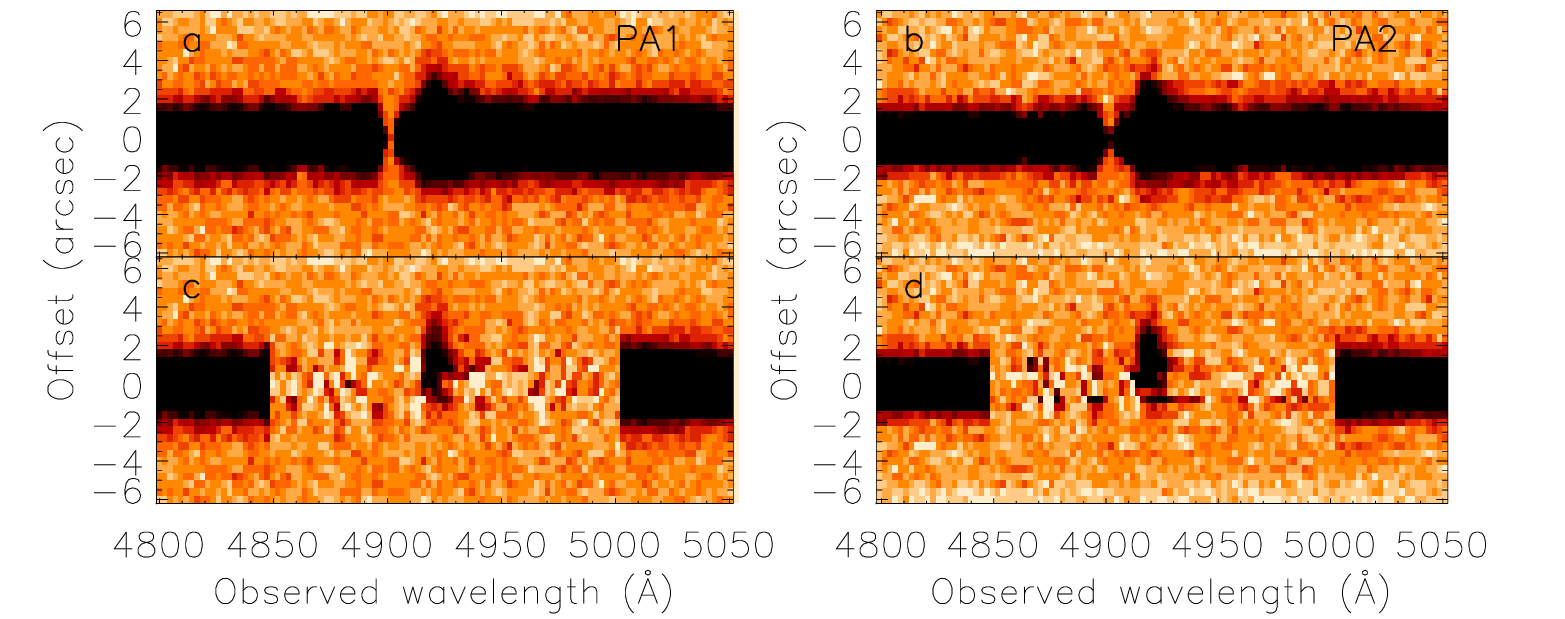,width=13cm}
\epsfig{file=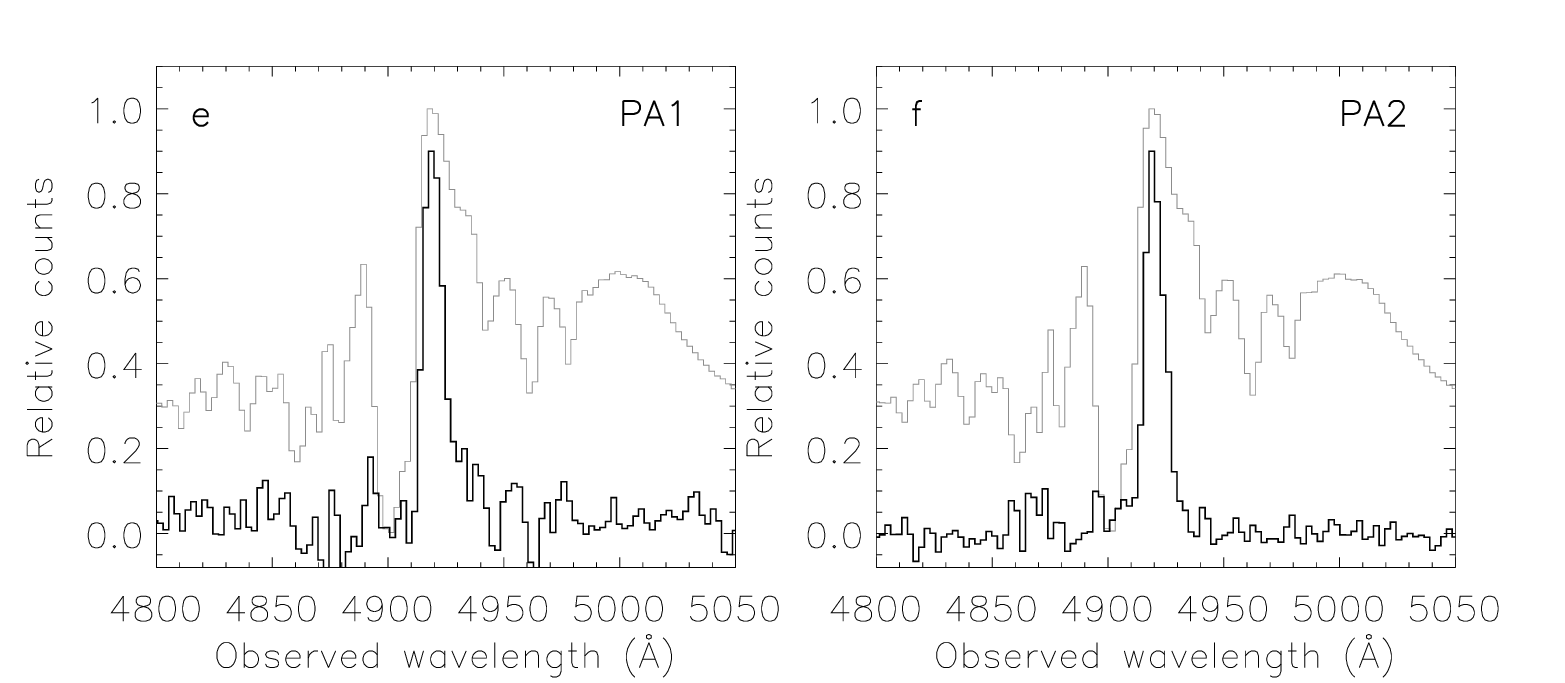,width=13cm}
\caption{Spectra at PA1 ({\it left column}) and PA2 ({\it right
  column}). {\bf a) -- b)} The 2D quasar spectrum. {\bf c) -- d)}
  The 2D quasar spectrum has been subtracted between $4850$ {\AA} and
  $5000$ {\AA} revealing the extended {\Lya} emission. The zero point
  of the spatial axis is set to the centroid of the QSO. Note the large
  residuals from the quasar SPSF subtraction, which are due to the
  large shot noise near the centroid. {\bf e) -- f)} The spectrum of
  the QSO {\Lya} emission line ({\it light grey curve}) and the spatially
  averaged spectrum of the extended {\Lya} emission ({\it black
  curve}). The maximum flux of the extended {\Lya} emission is scaled
  to $90\%$ of the maximum QSO flux. The total line flux in extended
  {\Lya} is around $1\%$ of the flux from the QSO {\Lya} line. The
  overall flux scale in this plot is arbitrary.}\label{fig:spectrum2d}
\end{figure*}

The science frames were coadded and the quasar spectrum was optimally
extracted using the code described in M{\o}ller (2000). After spectral
point-spread-function (SPSF) fitting and removal of the QSO, the
extended {\Lya} emission of S6 was clearly visible at both PA1 and PA2
(see Fig.~\ref{fig:spectrum2d}).

PA1 was aligned with the two galaxies g1 and g2 (see
Fig.~\ref{fig:field}) and centred on the QSO, so in this case we had to
decompose the spectrum into its individual components. We employed an
iterative procedure to separate the contributions from g1 and S6 to the
total quasar flux: {\it i)} Extract and remove the QSO spectrum. {\it
ii)} Extract and remove the spectrum of g1 and S6. The procedure
converged to a stable solution after three iterations. For the PA1
observations using grism G600R the seeing conditions were slightly
better than for those using G600B (see Table~\ref{tab:exp}), so
the projected distances of g1 and g2 were enough to bring them outside
the QSO point-spread function (PSF) and we could make a normal
extraction of the quasar spectrum.

\begin{figure*}[tbp]
\centering\epsfig{file=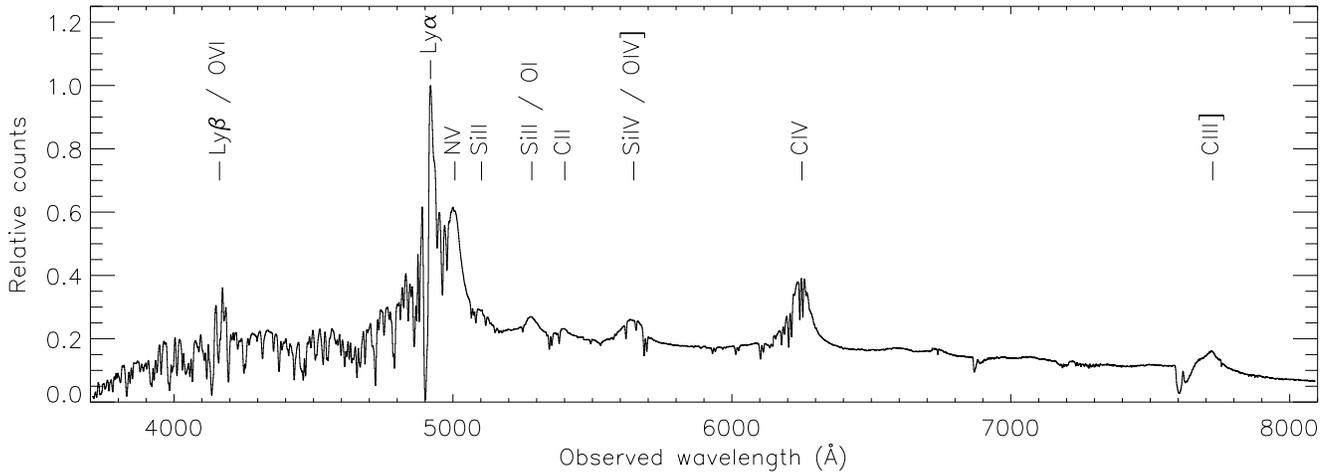,width=18cm}
\caption{ Extracted spectrum of \object{Q1205-30}. The most prominent emission
  lines are indicated. The spectrum has not been extinction
    corrected nor flux calibrated. }\label{fig:QSO}
\end{figure*}

\begin{figure*}[tbp]
\centering
\epsfig{file=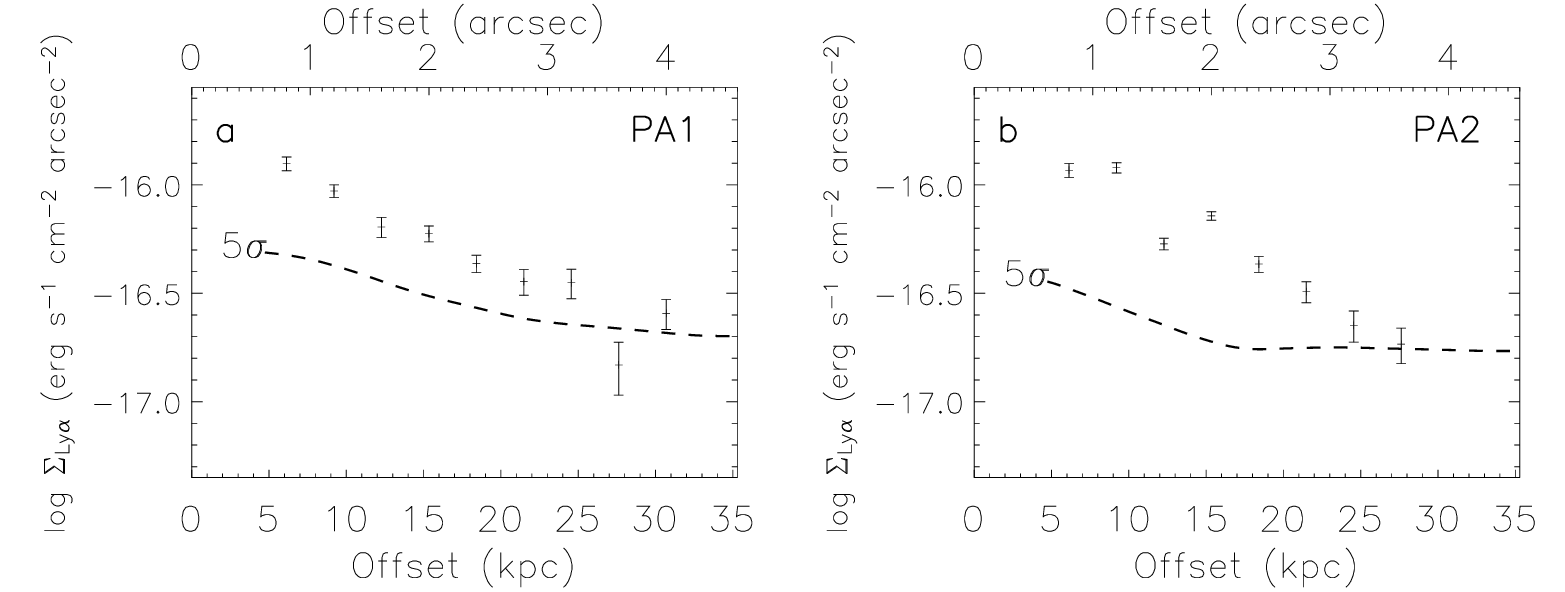,width=13cm}
\epsfig{file=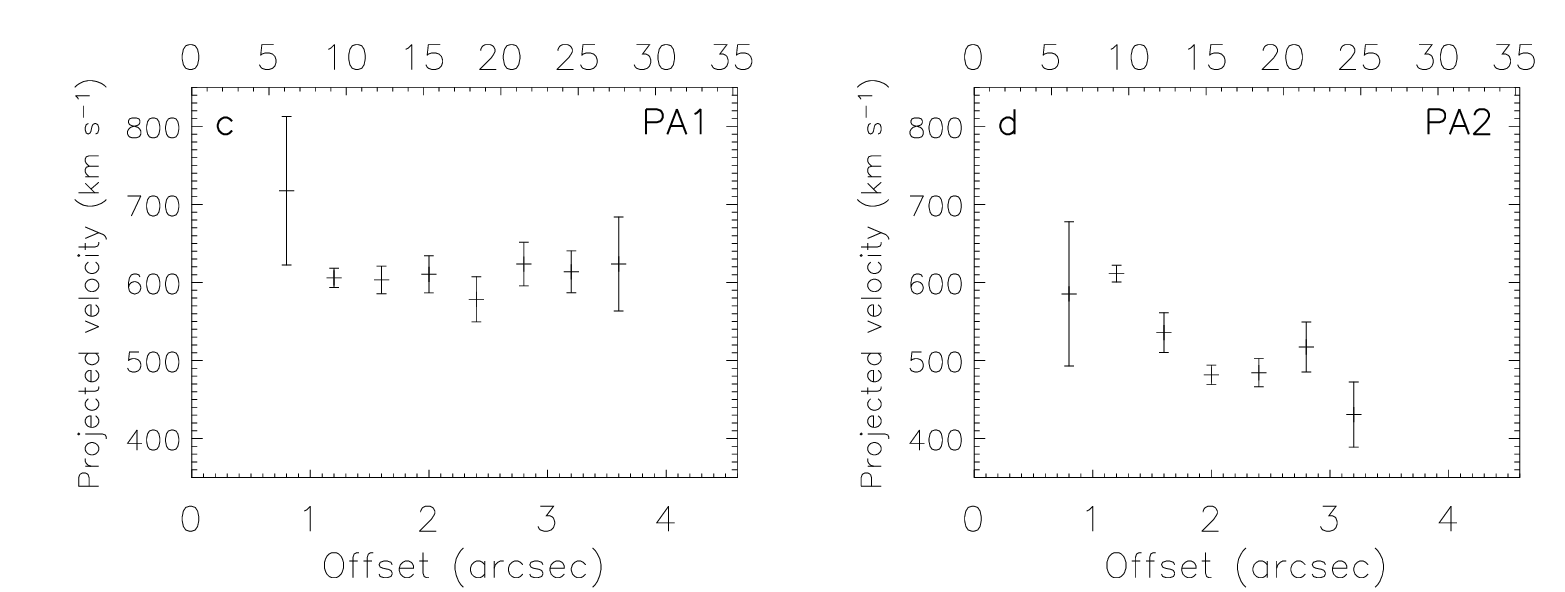,width=13cm}
\caption{Surface brightness and velocity profiles of the extended
{\Lya} emission based on spectra at PA1 ({\it left column}) and PA2
({\it right column}). {\bf a) -- b)} Surface brightness profile.
The error bars are based on photon statistics only. Systematic errors
due to the flux calibration are not considered. The thick dashed line
shows our $5\sigma$ detection limit. {\bf c) -- d)} Velocity profile
of the extended {\Lya} emission measured relative to the systemic
redshift of the quasar (see Sect.~\ref{sec:systemicz}). The
uncertainties are statistical uncertainties on the position of the
maximum of a fitted Gauss curve (Landman et al.~1982). In both PAs the
most distant detection was too faint to allow a secure determination of
the velocity. A rest-frame velocity of $600$ km s$^{-1}$ relative to
the systemic quasar redshift (see Sect.~\ref{sec:systemicz})
corresponds to a redshift $z=3.049$.}\label{fig:result}
\end{figure*}

In the PA2 spectra we had only the QSO and S6 on the slit. In order to
make sure that the extended flux of S6 was not modifying the quasar
SPSF, we here used an option in the code which allowed us to exclude a
wavelength region from the construction of the quasar SPSF. In the
region from 4902~{\AA} to 4987~{\AA} the code therefore only fitted and
extracted the QSO spectrum, it did not update the SPSF lookup table (for
details see M{\o}ller 2000). For the general problem of decomposing a
2D spectrum of two superimposed objects there is a certain degeneracy
of solutions. One may choose to assign the maximum amount of flux to
one object, to the other object, or to aim for somewhere in
between. In this case we decided to opt for a solution in between,
which avoids digging a hole in the extended {\Lya} emission near the QSO
centroid. The degeneracy has no consequence at distances larger than 1
arcsec from the QSO, but closer to the QSO the {\Lya} surface
brightness of S6 is very uncertain.

The output from the code is the optimally extracted 1D and 2D spectrum
of the QSO as well as the 2D spectrum of the extended {\Lya}
emission. The 2D spectrum of the extended {\Lya} emission and the 1D
spectrum of the QSO (with cosmic ray hits removed) are shown in
Fig.~\ref{fig:spectrum2d}a-\ref{fig:spectrum2d}d and
Fig.~\ref{fig:QSO}, respectively. For comparison we show a zoom of the
{\Lya} emission of the quasar and the spatially averaged extended {\Lya}
emission in Fig.~\ref{fig:spectrum2d}e-\ref{fig:spectrum2d}f.

\subsection{Wavelength and flux calibrations}\label{sec:cal}

The spectra in Fig.~\ref{fig:spectrum2d} were wavelength calibrated
using the {\it dispcor} task in IRAF\footnote{IRAF is distributed by
the National Optical Astronomy Observatories, which are operated by the
Association of Universities for Research in Astronomy, Inc., under
cooperative agreement with the National Science Foundation.}. The RMS
of the deviations from a 4.$^{\rm th}$ order Chebychev polynomial fit
to 12-19 lines were $0.6$~{\AA} for the G600B spectra and
$0.08-0.16$~{\AA} for 28-33 lines in the G600R spectra.

The flux calibration was done as follows. First we estimated the
continuum of the QSO away from the emission lines. We divided the
spectrum by the continuum, obtaining a flat spectrum, and forced it
onto an arbitrary power-law. Measuring the Bessel $B-I$ colour on the
resulting spectrum and comparing to the observed $B-I$ colour (Paper I)
we calculated the slope of the power-law which would bring the two in
agreement ($\alpha=0.65$). We then forced the spectrum onto this power
and normalized it to agree with our photometric measurements. The
advantage of this procedure is that to the first several orders all
absolute and differential slit losses as well as atmospheric absorption
are automatically taken into account for all point source objects on
the slit. We estimate the absolute flux calibration to be correct
to within $10\%$ while the relative is better. This applies to
point sources, but not to extended sources for which there are additional
slit losses. We present surface brightness profiles and 
rest-frame velocity curves of the extended {\Lya} emission (relative to
the systemic redshift of the quasar, see Sect.~\ref{sec:systemicz}) in
Fig.~\ref{fig:result}.

\section{Results}\label{sec:q1205:3}

\subsection{Redshifts of the galaxies g1 and g2}
The primary purpose of this study is to clarify the nature of the
extended {\Lya} emission at $z\approx 3.04$. As pointed out in Paper I the
galaxies g1 and in particular g2 may have a lensing effect, enhancing
and stretching the {\Lya} patch. Therefore we first set out to
determine their redshifts.

\begin{figure}[tbp]
\centering\epsfig{file=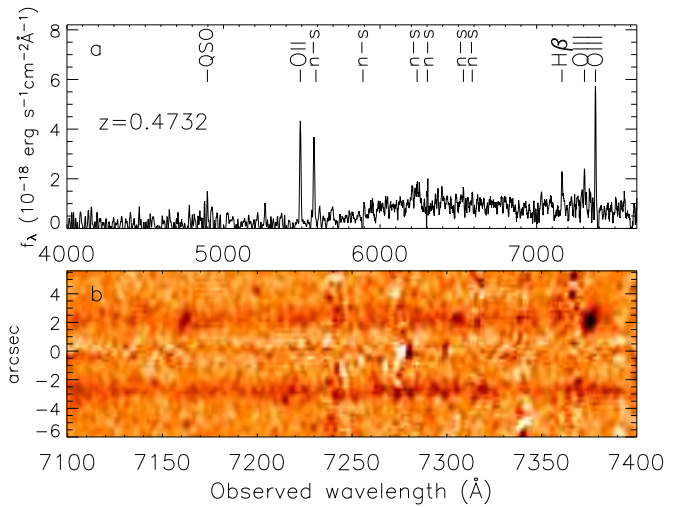,width=8.8cm}
\caption{{ {\bf a)} Spectrum of the blue galaxy g1 smoothed
with a $7$~{\AA} wide boxcar filter. The identified lines are
indicated, and strong residuals from night sky lines are marked
``n--s''. {\bf b)} 2D spectrum of the QSO residuals (at
$0$~arcsec), the blue galaxy g1 (at $2.12$~arcsec), and the red
galaxy g2 (at $-2.77$~arcsec). The rotation profile of g1 is clearly
visible in the emission lines H$\beta$ at 7162~{\AA} and \ion{O}{iii} at
7376~{\AA}. }}\label{fig:bg}
\end{figure}

\begin{figure}[tbp]
\centering\epsfig{file=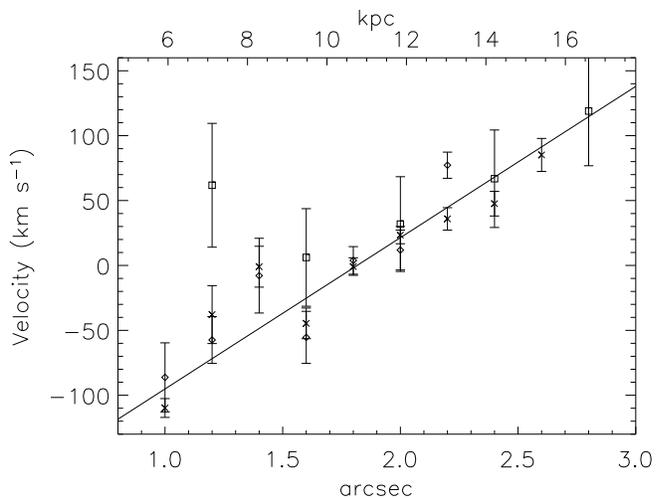,width=8.8cm}
\caption{{ Rotation profile of galaxy g1 mapped out from the
\ion{O}{ii} $\lambda 3727$ ($\square$), H$\beta~\lambda 4862$
($\diamond$) and \ion{O}{iii} $\lambda 5007$ ($\times$) lines. The
straight line is the best combined fit to all velocity profiles. The
$x$-axis denotes the distance from the QSO in arcsec (bottom) and kpc
(top), whereas the $y$-axis is the velocity relative to the redshift of
\ion{O}{iii} $\lambda 5007$ ($z=0.47311$). }}\label{fig:g1rot}
\end{figure}

We identify four emission lines in the spectrum of the blue galaxy
g1. The AB-magnitudes of g1 are $B(AB) = 24.8$, $I(AB) = 23.3$,
$n(AB) = 24.1$ (Paper I). Disregarding the weak and noisy \ion{O}{iii}
$\lambda 4960$ line, we derive a mean redshift of $z =
0.4732 \pm 0.00011$ (Table~\ref{tab:g1}). In the 2D spectrum
(Fig.~\ref{fig:bg}) we see along the slit a tilt of the emission
lines caused by the rotation of g1. The $\lambda 4960$ line is too
weak, but for the three remaining lines \ion{O}{ii} $\lambda 3727$,
H$\beta$, and \ion{O}{iii} $\lambda 5007$, we mapped out their rotation
profiles. The three profiles are identical within the errors, so we
combined the profiles and performed a joint fit (see
Fig.~\ref{fig:g1rot}). From linear extrapolation of the rotation
profile down to the position of the QSO ($b=0$~kpc), we find that g1
would cross the QSO spectrum at $-212\pm 20$~km~s$^{-1}$ corresponding
to $z=0.4722\pm 0.0001$. Assuming instead a flattened rotation curve
after the observed point closest to the QSO, we find that g1 would
cross the QSO spectrum at $-100 \pm 10$ km s$^{-1}$ corresponding to
$z=0.4727\pm  0.0001$. We shall return to a search for absorption at
this redshift in Sect.~\ref{sec:abslines}.

\begin{table}[tbp]
\caption{{ Emission lines of g1.}}\label{tab:g1}
\centering\begin{tabular}{c c c c }
\hline\hline
 Line & $\lambda_{\rm vac}$ & $z_{\rm em}$ \\
 & {\scriptsize (\AA)} &  \\\hline
 \ion{O}{ii}  $\lambda 3727$ & $5492.21$ & $0.4733(5)$ \\
 H$\beta$     $\lambda 4862$ & $7163.67$ & $0.4731(9)$ \\
 \ion{O}{iii} $\lambda 4960$ & $7308.07$ & $0.4733(2)$ \\
 \ion{O}{iii} $\lambda 5007$ & $7377.66$ & $0.4731(1)$ \\\hline
\end{tabular}\\
\end{table}

\begin{figure}[bp]
\centering\epsfig{file=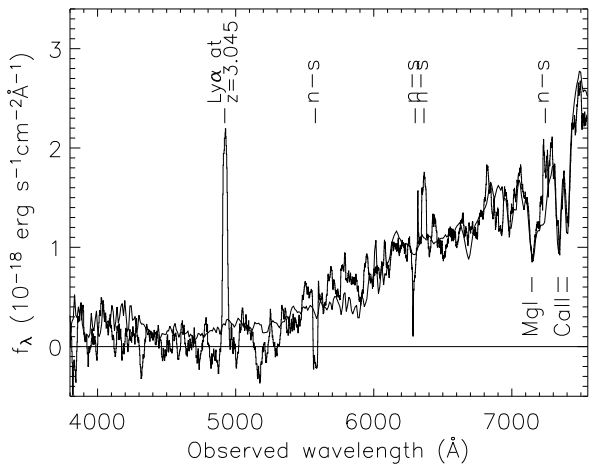,width=8.8cm}
\caption{{ Spectrum of g2 (thin line) smoothed with a $35$~{\AA} wide
boxcar filter, and template of an elliptical galaxy (thick line) at
$z=0.865$ (Kinney et al.~1996). Strong residuals from night sky lines
are marked ``n--s''.}}\label{fig:g2}
\end{figure}

The red galaxy g2 has no emission but absorption lines (see
Fig.~\ref{fig:g2}). The AB-magnitudes of g2 are $B(AB) > 26.5$,
$I(AB) = 22.22$, $n(AB) > 24.8 $ (Paper I). It was suggested in Paper I
that g2 most likely is a normal elliptical galaxy at $z\gtrsim 0.5$.
Minimum-$\chi^2$ fitting to redshifted template elliptical galaxy spectra
(Kinney et al.~1996) gave a best fit redshift of $z=0.865 \pm 0.003$,
confirming the prediction of Paper I. The observed spectrum and the
best-fit template are shown in Fig.~\ref{fig:g2}. The observed
I-band magnitude of g2 corresponds roughly to an absolute B-band
magnitude of $M_B = -22.1$, which is $0.7$ mag brighter than $M_B^*$
for field galaxies in the redshift interval $0.75 < z < 1.0$ (Cross
et al.~2004)

\subsection{The systemic redshift of the QSO}\label{sec:systemicz}
The quasar spectrum contains a Lyman-limit system (LLS) very close to
the redshift of the quasar (Lanzetta et al.~1991). In order to
establish the relation between the QSO, the LLS and the extended {\Lya}
emission we need to determine the precise redshifts of all three. For
QSOs this is not entirely straight forward. It is well known that
high-ionization lines are blueshifted with respect to the QSO systemic
redshift by typically several hundred km~s$^{-1}$ (e.g.~Tytler {\&} Fan
1992). For low-ionization lines the blueshifts are known to be small or
zero.

We measure equivalent widths, line fluxes and vacuum-corrected
wavelengths for eight both high- and low-ionization emission lines in
the quasar spectrum after removing absorption lines from intervening
systems. The results are presented in Table~\ref{tab:lines}. Because
the LLS redshift is very close to the QSO redshift, the quasar {\Lya}
emission line is heavily absorbed, so the observed centre wavelength is
dependent on how the line is reconstructed, resulting in an uncertainty
of $20$~{\AA}. We calculate the systemic redshift of the quasar
using the subset of the lines listed in Table~\ref{tab:lines} which
also appear with blueshift-adjusted rest-frame wavelengths in Tytler \&
Fan (1992). The five lines used are \ion{N}{v}, \ion{Si}{ii} /
\ion{O}{i}, \ion{Si}{iv} / \ion{O}{iv}], \ion{C}{iv}, and
\ion{C}{iii}]. The inverse-variance weighted average of these lines
provides a final $z_{\rm QSO} = 3.041 \pm 0.001$.

\begin{table*}[htbp]
\caption{{ Measured characteristics of eight QSO emission lines. The
columns show the line identification, the observed wavelength in
vacuum, the equivalent widths in observers frame and rest frame, the
line flux, the blueshift-adjusted rest wavelengths from Tytler \& Fan
(1992), and the corresponding redshift. The error associated with the
placement of the continuum is included in the errors on
$W_\mathrm{obs}$, $W_\mathrm{rest}$ and the line
flux. The {\Lya} line is heavily absorbed, so the redshift derived
from this line (shown in square brackets) is not used in the
calculation of the systemic QSO redshift. }}\label{tab:lines}
\centering\begin{tabular}{l c c c c c c c}
\hline\hline Line & $\lambda_{\rm rest}$$^a$ & $\lambda_\mathrm{vac}$ &
$W_\mathrm{obs}$ & $W_\mathrm{rest}$ & Line flux & $\lambda_{\rm TF}$ &
$z_{\rm TF}$ \\ & {\scriptsize (\AA)} &{\scriptsize (\AA)} & {\scriptsize
(\AA) } & {\scriptsize (\AA)} &{\scriptsize
$(10^{-16}$~erg~s$^{-1}$~cm$^{-2})$} & {\scriptsize(\AA)} & \\\hline
Ly$\alpha$ & $1215.67$ & $4918  \pm 20$  & $281 \pm 15$ & $69 \pm 4$  &
$541 \pm 33$ &$1214.97\pm0.07$& [$3.048\pm 0.016$]\\ \ion{N}{v}  &
$1240.15$ & $5005 \pm 5$ & $96 \pm 4$ & $24 \pm 1.0$  & $183\pm27$
&$1239.16\pm0.28$&$3.039\pm 0.004$\\ \ion{Si}{ii} & $1263.31$ & $5101
\pm 7$  &  $6 \pm 2$ & $1.5 \pm 0.5$ &  $10\pm3$ &---&---\\
\ion{Si}{ii} / \ion{O}{i}  & $1305.57$ & $5281 \pm 7$ & $15 \pm 2$ &
$3.8\pm0.6$ & $22.7 \pm 1.5$&$1304.24\pm0.30$& $3.049\pm 0.005$\\
\ion{C}{ii}  & $1335.31$ & $5400 \pm 3$ &  $9 \pm 2$ & $2.2 \pm 0.5$ &
$11\pm2$ &---&---\\  \ion{Si}{iv} / \ion{O}{iv}] & $1399.55$ & $5647
\pm 3$ & $62 \pm 9$ & $15 \pm 2$  & $81\pm4$&$1398.62\pm0.19$&
$3.037\pm 0.002$\\ \ion{C}{iv}  & $1549.05$ & $6249 \pm 5$ & $180 \pm
15$ & $44 \pm 4$  & $195\pm11$ &$1547.46\pm0.04$&$3.038\pm 0.003$\\
\ion{C}{iii}] & $1908.73$ & $7722 \pm 7$ & $100 \pm 12$ & $25 \pm 3$ &
$94\pm9$ &$1906.53\pm0.09$&$3.050\pm 0.004$\\ \hline
\end{tabular}
\\ {\scriptsize $^a$ Taken from Wilkes (2000).}
\end{table*}

\subsection{Absorption line systems}\label{sec:abssystems}
\begin{figure*}[tbp]
\centering\epsfig{file=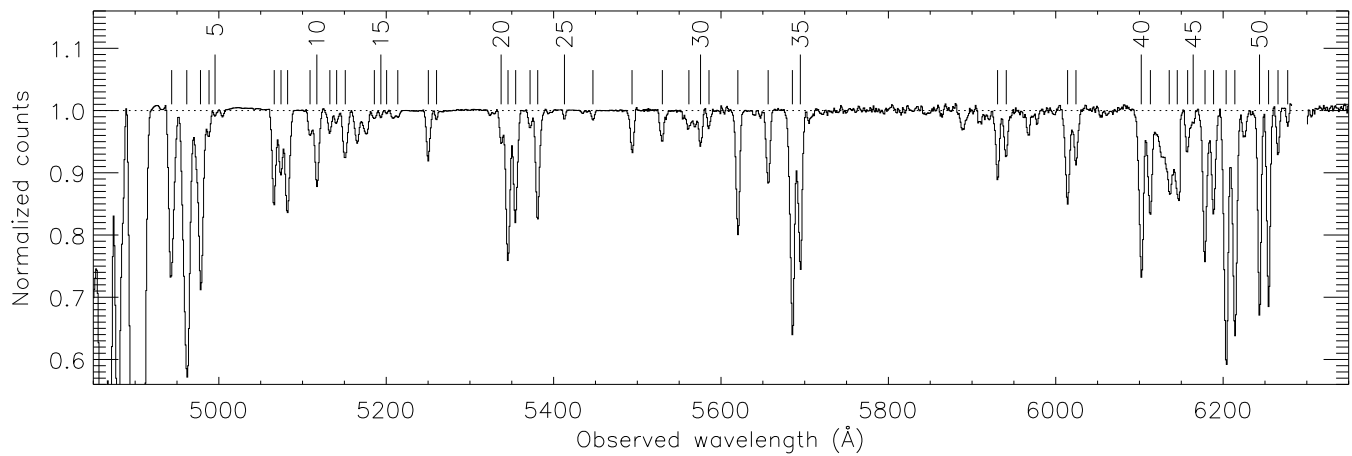,width=18cm}
\caption{Zoom on the normalized spectrum of \object{Q1205-30} between
the {\Lya} and \ion{C}{iv} QSO emission lines. The numbering of the
absorption lines correspond to the numbering in
Table~\ref{tab:abs}. }\label{fig:abslines}
\end{figure*}

In the quasar spectrum we identify 12 \ion{C}{iv} absorption systems
between the {\Lya} and \ion{C}{iv} emission lines. In
Fig.~\ref{fig:abslines} we plot this section of the QSO spectrum
normalized to the continuum. The line identifications are presented in
Table~\ref{tab:abs}, and the redshifts and identified lines for the 12
systems are summarized in Table~\ref{tab:systems}. We expect to detect
$5.2\pm 1$ \ion{C}{iv} absorbers per unit redshift (Fig.~5 in Sargent et
al.~1988). We find $13.9$ systems per unit redshift between the {\Lya}
($z=2.17$) and \ion{C}{iv} ($z=3.04$) lines of the QSO. Applying the same
selection criteria as the S2 sample of Sargent et al., namely rest
equivalent widths $W_{\rm rest} > 0.15$ {\AA} for both lines in the
\ion{C}{iv} doublet, a rest-frame velocity relative to the QSO $v <
-5000$ km s$^{-1}$ (in this case corresponding to $z_{\rm abs} <
2.974$), and grouping systems separated by less than $1000$ km
s$^{-1}$, we are left with $5.8\pm 2.4$ systems per unit
redshift. Using these criteria Sargent et al.~find $2.2^{+0.7}_{-0.5}$
systems per unit redshift at $z = 2 - 3$ (Fig.~6 in Sargent et
al.~1988). Our result is higher than normally observed, but still
marginally consistent with the Sargent et al.~study.

\begin{table}[tbp]
\caption{{ Identified absorption lines redwards of \Lya. The columns
show the line number, the observed wavelength in vacuum, the observed
equivalent width, the inferred redshift and the absorption line and
system identification. The error bars are $1\sigma$ errors due to
read-out noise and photon statistics only. The error associated with
the position of the continuum is not considered. The line numbers refer
to the numbering in Fig.~\ref{fig:abslines}. }}\label{tab:abs}
\centering\begin{tabular} { >{\scriptsize}c >{\scriptsize}c
>{\scriptsize}c >{\scriptsize}c >{\scriptsize}l >{\scriptsize}c}
\hline\hline \# &$\lambda_{\rm vac}$ & $W_{\rm obs}$ & $z_{\rm abs}$ &
\multicolumn{2}{l}{\scriptsize Identification} \\ & {(\AA)}    & {(\AA)}
&          &             &  \\\hline 1 & $4945.0$ & $1.92\pm 0.02$ &
$2.9917$ & \ion{N}{v}    & I \\ 2 & $4963.0$ & $3.82\pm0.02$&$2.9934 /
3.0063$&\ion{N}{v}&I / J \\ 3 & $4979.4$ & $2.16\pm 0.02$ & $3.0066$ &
\ion{N}{v}    & J \\ 4 & $4989.6$ & $0.18\pm 0.02$ & $2.8318$ &
\ion{O}{i}    & E \\ 5 & $4996.8$ & $0.03\pm 0.02$ & $2.8308$ &
\ion{Si}{ii}  & E \\ 6 & $5067.6$ & $0.83\pm 0.03$ & $2.2732$ &
\ion{C}{iv}   & A \\ 7 & $5075.7$ & $0.57\pm 0.02$ & $2.2730$ &
\ion{C}{iv}   & A \\ 8 & $5083.5$ & $1.02\pm 0.03$ & $3.0332$ &
\ion{Si}{ii}  & K \\ 9 & $5110.4$ & $0.20\pm 0.02$ & $2.8293$ &
\ion{C}{ii}   & E \\ 10 & $5118.5$ & $0.73\pm 0.03$ & $2.6725$ &
\ion{Si}{iv}  & D \\ 11 & $5134.1$ & $0.20\pm 0.03$ & $2.9427$ &
\ion{O}{i}    & G \\ 12 & $5142.2$ & $0.10\pm 0.02$ & $2.9423$ &
\ion{Si}{ii}  & G \\ 13 & $5152.5$ & $0.50\pm 0.04$ & $2.6731$ &
\ion{Si}{iv}  & D \\ 14 & $5187.2$ & $0.07\pm 0.03$ & $2.8869$ &
\ion{C}{ii}   & F \\ 15 & $5195.0$ & $0.04\pm 0.02$ & $2.9895$ &
\ion{O}{i}    & I \\ 16 & $5201.9$ & $0.01\pm 0.02$ & $2.9880$ &
\ion{Si}{ii}  & I \\ 17 & $5215.3$ & $0.06\pm 0.03$ & $3.0051$ &
\ion{O}{i}    & J \\ 18 & $5251.8$ & $0.40\pm 0.03$ & $3.0331$ &
\ion{O}{i}    & K \\ 19 & $5261.6$ & $0.05\pm0.03$&$2.9426$ / $3.0338$
&\ion{C}{ii} / \ion{Si}{ii}  & G / K\\ 20 & $5338.7$ & $0.23\pm 0.03$ &
$2.8305$ & \ion{Si}{iv}  & E \\ 21 & $5347.0$ & $1.48\pm 0.03$ &
$2.4537$ & \ion{C}{iv}   & B \\ 22 & $5356.1$ & $0.97\pm 0.03$ &
$2.4538$ & \ion{C}{iv}   & B \\ 23 & $5373.3$ & $0.13\pm 0.03$ &
$2.8305$ & \ion{Si}{iv}  & E \\ 24 & $5382.5$ & $0.97\pm 0.03$ &
$3.0332$ & \ion{C}{ii}   & K \\ 25 & $5414.5$ & $0.04\pm 0.02$ &
$2.8848$ & \ion{Si}{iv}  & F \\ 26 & $5448.5$ & $0.07\pm 0.03$ &
$2.8841$ & \ion{Si}{iv}  & F \\ 27 & $5495.2$ & $0.34\pm 0.03$ &
$2.9427$ & \ion{Si}{iv}  & G \\ 28 & $5531.5$ & $0.29\pm 0.03$ &
$2.9433$ & \ion{Si}{iv}  & G \\ 29 & $5563.2$ & $0.22\pm 0.03$ &
$2.9915$ & \ion{Si}{iv}  & I \\ 30 & $5577.1$ & $0.28\pm 0.05$ &
$2.6023$ & \ion{C}{iv}   & C \\ 31 & $5587.2$ & $0.13\pm 0.02$ &
$2.6028$ & \ion{C}{iv}   & C \\ 32 & $5621.7$ & $1.05\pm 0.03$ &
$3.0335$ & \ion{Si}{iv}  & K \\ 33 & $5657.9$ & $0.63\pm 0.02$ &
$3.0334$ & \ion{Si}{iv}  & K \\ 34 & $5686.9$ & $2.29\pm 0.02$ &
$2.6732$ & \ion{C}{iv}   & D \\ 35 & $5696.4$ & $1.41\pm 0.02$ &
$2.6732$ & \ion{C}{iv}   & D \\ 36 & $5932.1$ & $0.58\pm 0.02$ &
$2.8316$ & \ion{C}{iv}   & E \\ 37 & $5942.5$ & $0.43\pm 0.03$ &
$2.8320$ & \ion{C}{iv}   & E \\ 38 & $6015.8$ & $0.92\pm 0.02$ &
$2.8857$ & \ion{C}{iv}   & F \\ 39 & $6025.8$ & $0.49\pm 0.02$ &
$2.8857$ & \ion{C}{iv}   & F \\ 40 & $6103.9$ & $1.69\pm 0.02$ &
$2.9425$ & \ion{C}{iv}   & G \\ 41 & $6114.8$ & $1.12\pm 0.02$ &
$2.9430$ & \ion{C}{iv}   & G \\ 42 & $6137.3$ & $1.19\pm 0.02$ &
$2.9641$ & \ion{C}{iv}   & H \\ 43 & $6146.9$ & $1.01\pm 0.33$ &
$2.9637$ & \ion{C}{iv}   & H \\ 44 & $6159.3$ & $0.36\pm 0.02$ &
$3.0344$ & \ion{Si}{ii}  & K \\ 45 & $6165.9$ & $0.10\pm 0.02$ &
$2.8334$ & \ion{Fe}{ii}  & E \\ 46 & $6180.0$ & $1.41\pm 0.02$ &
$2.9918$ & \ion{C}{iv}   & I \\ 47 & $6190.2$ & $0.97\pm 0.02$ &
$2.9917$ & \ion{C}{iv}   & I \\ 48 & $6205.3$ & $2.52\pm 0.02$ &
$3.0081$ & \ion{C}{iv}   & J \\ 49 & $6215.7$ & $2.20\pm 0.02$ &
$3.0081$ & \ion{C}{iv}   & J \\ 50 & $6245.2$ & $1.55\pm 0.01$ &
$3.0338$ & \ion{C}{iv}   & K \\ 51 & $6256.0$ & $1.47\pm 0.01$ &
$3.0341$ & \ion{C}{iv}   & K \\ 52 & $6267.3$ & $0.29\pm 0.01$ &
$3.0481$ & \ion{C}{iv}   & L \\ 53 & $6278.8$ & $0.08\pm 0.01$ &
$3.0488$ & \ion{C}{iv}   & L \\\hline
\end{tabular}
\end{table}

\begin{table}[tbp]
\caption{Overview of the absorption systems in the spectrum of
  \object{Q1205-30}.}
\label{tab:systems}
\centering\begin{tabular}{ c p{4.4cm} c }
\hline\hline
 System & Detected lines    & $z_{\rm abs}$ \\\hline
 A   & \ion{C}{iv} $\lambda\lambda 1548$, $1550$ & $2.2731 \pm 0.0001$ \\
 B   & \ion{C}{iv} $\lambda\lambda 1548$, $1550$ & $2.4538 \pm 0.0001$ \\
 C   & \ion{C}{iv} $\lambda\lambda 1548$, $1550$ & $2.6026 \pm 0.0004$ \\
 D   & \ion{Si}{iv} $\lambda\lambda 1393$, $1402$, \ion{C}{iv} $\lambda\lambda 1548$, $1550$ & $2.6730 \pm 0.0004$ \\
 E   & \ion{O}{i} $\lambda 1302$, \ion{Si}{ii} $\lambda 1304$, \ion{C}{ii} $\lambda 1334$, \ion{Si}{iv} $\lambda\lambda 1393$, $1402$, \ion{C}{iv} $\lambda\lambda 1548$, $1550$, \ion{Fe}{ii} $\lambda 1608$ & $2.8309 \pm 0.0009$ \\
 F   & \ion{C}{ii} $\lambda 1334$, \ion{Si}{iv} $\lambda\lambda 1393$, $1402$, \ion{C}{iv} $\lambda\lambda 1548$, $1550$ & $2.885 \pm 0.001$ \\
 G   & \ion{O}{i} $\lambda 1302$, \ion{Si}{ii} $\lambda 1304$,
 \ion{C}{ii} $\lambda 1334$, \ion{Si}{iv} $\lambda\lambda 1393$, $1402$, \ion{C}{iv} $\lambda\lambda 1548$, $1550$ & $2.9427 \pm 0.0003$ \\
 H   & \ion{C}{iv} $\lambda\lambda 1548$, $1550$ & $2.9639 \pm 0.0003$ \\
 I   & \ion{N}{v} $\lambda\lambda 1238$, $1242$, \ion{O}{i} $\lambda 1302$, \ion{Si}{ii} $\lambda 1304$, \ion{Si}{iv} $\lambda 1393^a$, \ion{C}{iv} $\lambda\lambda 1548$, $1550$ & $2.991 \pm 0.002$ \\
 J   & \ion{O}{i} $\lambda 1302$, \ion{N}{v} $\lambda\lambda 1238$, $1242$, \ion{C}{iv} $\lambda\lambda 1548$, $1550$ & $3.007 \pm 0.001$ \\
 K   & \ion{Si}{ii} $\lambda\lambda 1260$, $1304$, $1526$, \ion{O}{i} $\lambda 1302$, \ion{C}{ii} $\lambda 1334$, \ion{Si}{iv} $\lambda\lambda 1393$, $1402$, \ion{C}{iv} $\lambda\lambda 1548$, $1550$ & $3.0336 \pm 0.0004$ \\
 L   & \ion{C}{iv} $\lambda\lambda 1548$, $1550$ & $3.0485 \pm 0.0005$ \\\hline
\end{tabular}\\
{\scriptsize $^a$ \ion{Si}{iv} $\lambda 1403$ was too weak to be
detected.}
\end{table}

\begin{figure}[tbp]
\centering\epsfig{file=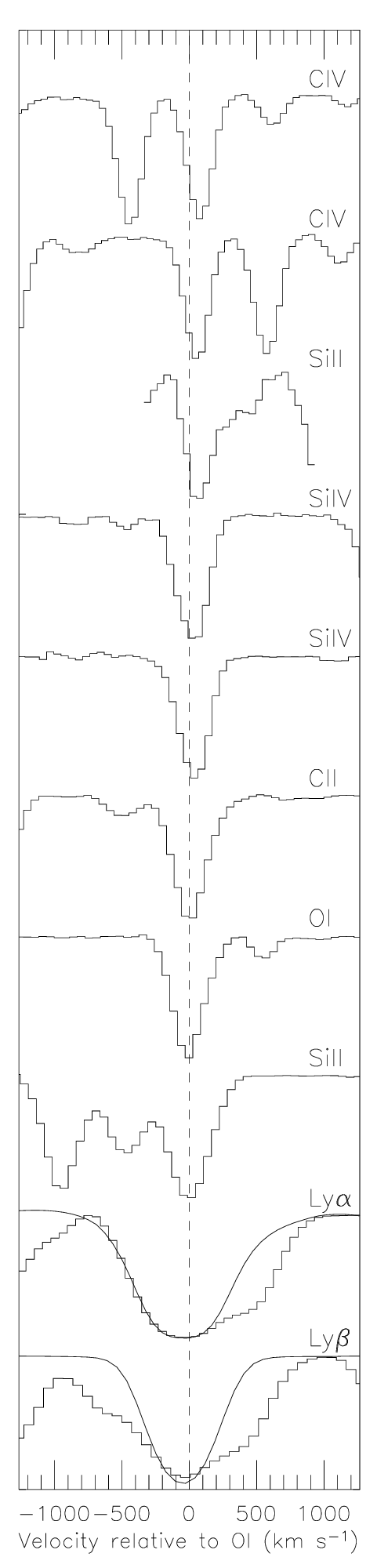,width=4.5cm}
\caption{Velocities of lines originating from absorption system K (the LLS)
  relative to $z_{\rm OI} = 3.0331$. The model plotted for {\Lya}
  and Ly$\beta$ has an \ion{H}{i} column density of $\log N_{\ion{H}{i}} =
  19.5$. The \ion{Si}{ii} $\lambda 1304$ transition is not included
  here due to the probable blend with \ion{C}{ii} $\lambda 1334$ at
  $z=2.9426$ (see Table \ref{tab:abs}).}\label{fig:abssystemJ}
\end{figure}

The LLS at $z\approx 3.034$, which was the original target (Paper I),
is identical to the absorption system K, for which both low- and
high-ionization lines are detected. This is in agreement with the
findings for other $z_{\rm abs} \approx z_{\rm em}$ absorbers (Savaglio
et al.~1994; Hamann 1997; M{\o}ller et al.~1998). By fitting line
profiles to the {\Lya} and Ly$\beta$ lines we find the \ion{H}{i}
column density to be in the interval $17\leq \log N_{\ion{H}{i}} \leq
19.9$. The best fit gives $\log N_{\ion{H}{i}} = 19.5$. In
Fig.~\ref{fig:abssystemJ} we plot the absorption lines originating from
system K in velocity space relative to the redshift obtained from the
\ion{O}{i} absorption line, $z_{\rm OI} = 3.0331 \pm 0.0002$. We notice
that high-ionization lines have systematically higher redshifts than
low-ionization lines. We use the redshift from the \ion{O}{i} line as
the redshift of the low-ionization region, and we take the redshift
obtained from the \ion{C}{iv} doublet, $z_{\rm CIV} = 3.0340 \pm
0.0002$, to be the redshift of the high-ionization region. We find that
the region of highly ionized elements moves with a velocity of $\sim
60$~km~s$^{-1}$ relative to the low-ionization region. Furthermore, we
notice an \ion{H}{i} absorption system in the red wing of both the
{\Lya} and Ly$\beta$ lines of the LLS (see
Fig.~\ref{fig:abssystemJ}). This system, which we will call system K1,
has a redshift of $3.039\pm 0.002$, and we find no associated
metal-lines.

We detect high-ionization \ion{N}{v} absorption for the systems I
and J. If the high degree of ionization is caused by the high UV flux
from the quasar, the systems must be located between the QSO and the
LLS at $z\approx 3.034$, since no UV photons pass through the LLS. The
redshifts of systems I and J are lower than that of the LLS (system K),
which suggests that systems I and J are high-velocity
clouds. Furthermore, the high degree of ionization suggests that they
are associated with the QSO. The apparent line-locking between
\ion{N}{v} $\lambda 1242$ of system I and \ion{N}{v} $\lambda 1238$ of
system J strengthens this hypothesis. The line-locking effect is
thought to occur in clouds driven by radiation pressure and accelerated
via absorption until the wavelength of the feature falls in the shadow
of another line in a neighbouring cloud (Vilkoviskij et al.~1999;
Srianand et al.~2002). The velocities of the absorption systems are
$-4050$ km s$^{-1}$ (system I) and $-2850$ km s$^{-1}$ (system J)
relative to the systemic redshift of the quasar.

\section{Foreground galaxies}\label{sec:abslines}

A classical way to search for galaxies at intermediate or high
redshifts has been to look for absorption systems in the spectra of
background QSOs (e.g.~Weymann et al.~1979). Studies of the galaxy
counterparts of \ion{Mg}{ii} absorption systems have concluded that the
galaxies responsible for the absorption are bright galaxies which can
be found at impact parameters of $\sim 100$~kpc (Bergeron {\&} Boiss\'e
1991, Guillemin {\&} Bergeron 1997). Using a sample of damped {\Lya}
absorbers (DLAs) Lanzetta et al.~(1995) found that most galaxy
counterparts of DLAs at $z\lesssim 1$ have gaseous haloes extending over
$\sim 230$~kpc. This is in contrast to studies at high redshift
($z = 2 - 3.5$), where M{\o}ller et al.~(2002) have found that galaxy
counterparts of DLAs reside at impact parameters $\sim 10$~kpc. The
large impact parameters have been used to advocate a picture of
galaxies surrounded by a large, homogeneous gaseous envelope which is
responsible for the absorption, but doubt has arisen whether the
galaxies at large projected distances are the true absorbers (Yanny \&
York 1992; Vreeswijk et al.~2003; Jakobsson et al.~2004).

\begin{figure}[tbp]
\centering\epsfig{file=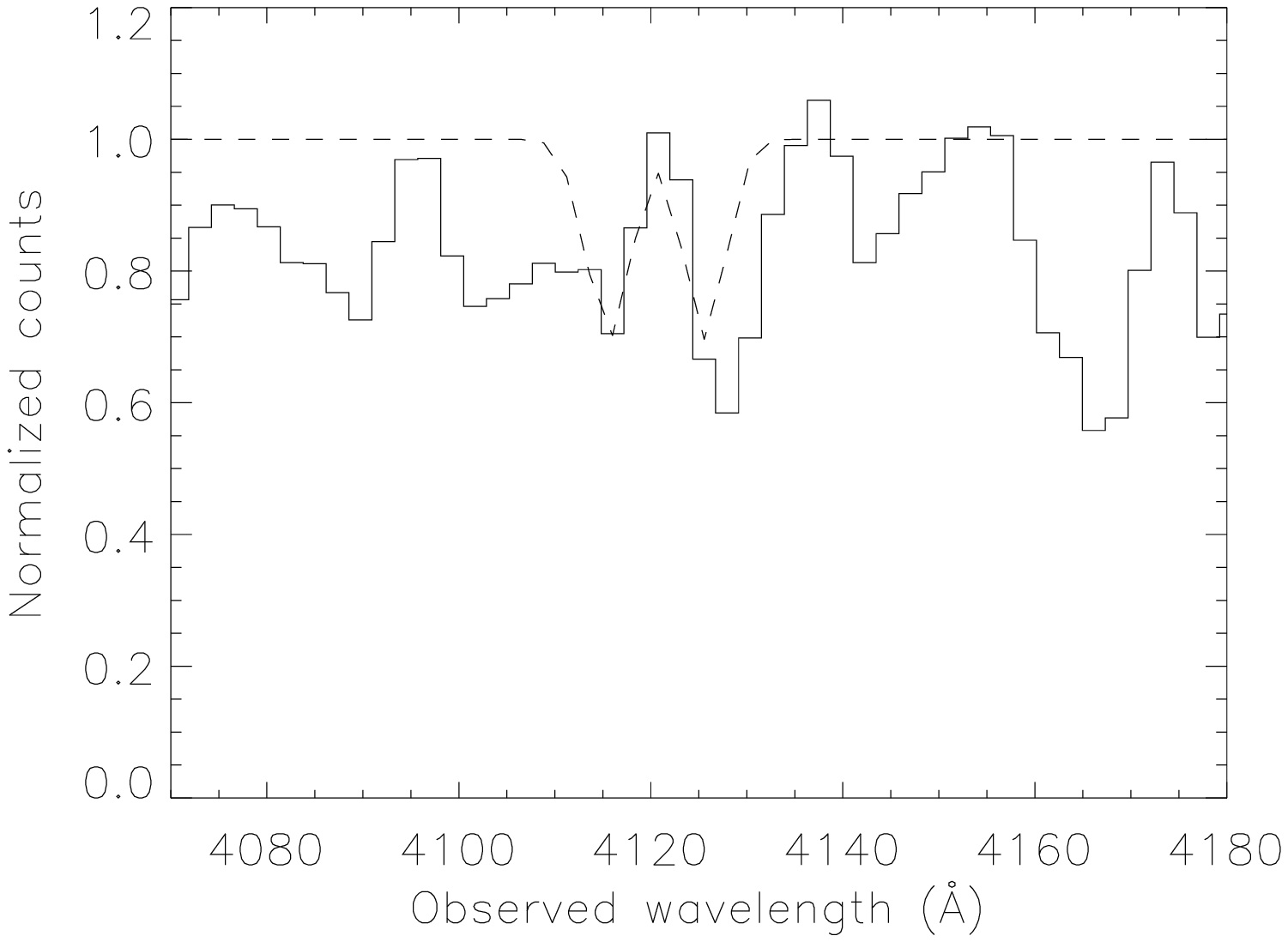,width=8.8cm}
\caption{Spectrum of \object{Q1205-30} normalized to the continuum. In the
  spectrum we have subtracted lines corresponding to \Lya, Ly$\beta$ and
  \ion{O}{vi} absorption from the systems listed in Table
  \ref{tab:systems}. The dashed line shows the model of a possible \ion{Mg}{ii}
  doublet from g1. }\label{fig:MgII}
\end{figure}

We have looked for absorption lines in the QSO spectrum due to the
foreground galaxies g1 and g2 located at impact parameters $12.6$
kpc and $21.3$ kpc, respectively. Considering the redshifts of g1 and g2,
the only lines covered by our spectrum are \ion{Ca}{ii} H and K,
\ion{Mg}{ii} $\lambda\lambda 2796$, $2804$, \ion{Mg}{i} $\lambda 2853$
for both galaxies, as well as \ion{Fe}{ii} $\lambda\lambda 2344$,
$2374$, $2383$, $2587$, $2600$ for g2. Lines in the Lyman-forest are
not well-constrained due to the large number of {\Lya} lines. We fitted
and removed lines corresponding to \Lya, Ly$\beta$ and \ion{O}{vi}
absorption from the \ion{C}{iv} systems listed in
Table~\ref{tab:systems}. In the residuals we identify two absorption
lines (see Fig.~\ref{fig:MgII}) which may be due to \ion{Mg}{ii} at
$z=0.4722$. Alternatively, they could be two Lyman-forest lines. In
Table~\ref{tab:uplim} we list the $2\sigma$ upper limits on the
equivalent widths of the other lines.

We find no absorption lines in the QSO spectrum outside the
Lyman-forest due to the two foreground galaxies g1 and g2 down to a
$2\sigma$ upper limit on the equivalent widths of $\sim 90$~m{\AA}
(rest frame), but we cannot exclude that g1 is a \ion{Mg}{ii} absorber
with equivalent width less than $\sim 1$~{\AA}, as the lines would fall
in the Lyman forest. Assuming that the g1 \ion{Mg}{ii} doublet
candidate is correctly identified, we cannot exclude that g1 is a DLA
(see Fig.~24 in Rao \& Turnshek 2000). Conversely, it is unlikely that
g2 is a DLA system (Fig.~25 and 26 in Rao \& Turnshek 2000).

Old elliptical galaxies have recently been found up to a redshift
$z=1.9$ (Cimatti et al.~2004), but not much is known about the gas
content in elliptical galaxies at these early times. The
combination of a small impact parameter (corresponding to $21.3$~kpc)
and strict upper limits on absorption for the elliptical galaxy g2
indicates that it has no significant gas column density at these
radii. The redshift $z=0.865$ of g2 corresponds to a lookback time of
$7.1$~Gyr.

\begin{table}[tbp]
\caption{Absorption line search in the QSO spectrum due to g1 and g2. The
 quoted upper limits are $2\sigma$ limits on the equivalent widths. The
 error associated with the placement of the continuum is taken into
 account for lines not in the Lyman-forest ($\lambda_{\rm obs} >
 4900$~\AA).}\label{tab:uplim}
\centering\begin{tabular} {l l c c c }
\hline\hline
 & Ion & $\lambda_{\rm rest}$ & $\lambda_{\rm expected}$ & $W_{\rm rest}$ \\
 &            & (\AA)      & (\AA)      & (\AA)\\\hline
{\bf g1}
 & \ion{Mg}{ii} & $2796.4$ & $4116.8$ & $1.2^a$ \\
 & \ion{Mg}{ii} & $2803.5$ & $4127.4$ & $1.1^a$ \\
 & \ion{Mg}{i}  & $2853.0$ & $4200.1$ & $< 0.20$ \\
 & \ion{Ca}{ii} & $3934.7$ & $5792.8$ & $< 0.095$ \\
 & \ion{Ca}{ii} & $3969.6$ & $5844.0$ & $< 0.068$ \\\hline
{\bf g2}
 & \ion{Fe}{ii} & $2344.2$ & $4372.0$ & $< 0.4$ \\
 & \ion{Fe}{ii} & $2374.4$ & $4428.4$ & $< 0.4$ \\
 & \ion{Fe}{ii} & $2382.8$ & $4443.9$ & $< 0.5$ \\
 & \ion{Fe}{ii} & $2586.6$ & $4824.1$ & $< 0.5$ \\
 & \ion{Fe}{ii} & $2600.2$ & $4849.3$ & $< 0.6$ \\
 & \ion{Mg}{ii} & $2796.4$ & $5215.2$ & $< 0.097$ \\
 & \ion{Mg}{ii} & $2803.5$ & $5228.6$ & $< 0.107$ \\
 & \ion{Mg}{i}  & $2853.0$ & $5320.8$ & $< 0.059$ \\
 & \ion{Ca}{ii} & $3934.7$ & $7338.4$ & $< 0.089$ \\
 & \ion{Ca}{ii} & $3969.6$ & $7403.3$ & $< 0.073$ \\\hline
\end{tabular}\\
{\scriptsize $^a$ The line is in the Lyman-forest, so it may be
{\Lya} at an intermediate redshift. The identification is therefore not
secure.}
\end{table}

\subsection{Gravitational lensing}
The position of g2 could introduce a lensing effect on S6. Knowing the
true redshift of g2, we can repeat the calculation in Paper I of the
radius of its Einstein ring. For easy comparison to Paper I we assume
that g2 has a singular isothermal mass distribution with a velocity
dispersion $\sigma = 300$~km~s$^{-1}$, in which case the radius of the
Einstein ring is
\begin{align}
\theta_E = 4\pi\left(\frac{\sigma}{c}\right)^2 \frac{d_{LS}}{d_S} =
1.4~\text{arcsec}
\end{align}
(from the equation following Eq.~4.14 in Peacock 1999), where $d_{LS}$
is the lens-source angular distance and $d_S$ is the observer-source
angular distance. Gravitational lensing will introduce stretching and
distortion perpendicular to the radius vector from g2, most notably at
a distance of one Einstein radius. Emission appearing within one
Einstein radius of the centre of g2 (roughly corresponding to the
dotted circle in the lower right part of Fig.~\ref{fig:field}) is
expected to originate from the same small region. This stretching
effect would result in a flat velocity profile along PA1, which is seen
in Fig.~\ref{fig:result}c over a distance of up to $4$ arcsec from
the QSO centre towards g2.

The fact that we only detect one lensed image of the quasar constrains
the mass of g2 within a projected distance corresponding to the g2 --
quasar angular distance (2.77 arcsec). Following the calculations of Le
Brun et al.~(2000), we may calculate a model independent upper limit on
the projected mass $M^\Sigma_{\rm max}(< r_{\rm I})$ enclosed within a
radius $r_{\rm I} = d_L \theta_{\rm I} = 21.3$ kpc:
\begin{align}
M^\Sigma_{\rm max}(< r_{\rm I}) = \pi r_{\rm I}^2 \Sigma_{\rm crit} =
2.8\times 10^{12} M_\odot,
\end{align}
where $\Sigma_{\rm crit} = \frac{c^2}{4\pi G}\frac{d_S}{d_L d_{LS}}$ is
the critical surface density. This corresponds to an upper limit
to the radius of the Einstein ring of $\theta_E < 2.77$ arcsec.

Conversely, the following lower limit to the size of the Einstein ring
allows us to constrain from below the projected mass. From
Fig.~\ref{fig:result}c it is evident that the Einstein radius
is at least $0.8$ arcsec (the constant part of the velocity
profile between galaxy g2 at $\sim 2.8$ arcsec and out to the most distant
measurement at $3.6$ arcsec). Thus $\theta_E > \theta_{\rm min} = 0.8$
arcsec, and
\begin{align}
M(<r_{\rm I}) > M(<d_L \theta_E) > \frac{\theta_{\rm min}^2
c^2}{4G}\frac{d_Sd_L}{d_{LS}} = 2.3\times 10^{11} M_\odot.
\end{align}

Thus the projected mass of galaxy g2 is between $2.3\times 10^{11}
M_\odot$ and $2.8\times 10^{12} M_\odot$ within a radius of $2.77$
arcsec ($21.3$ kpc). This is consistent with the super-$M_B^*$ finding of
Sect.~\ref{sec:q1205:3}. In terms of velocity dispersion the range is
$230\text{ km s}^{-1} < \sigma < 420\text{ km s}^{-1}$, assuming a singular
isothermal mass distribution.

\section{A model of the extended {\Lya} emission}\label{sec:model}

In trying to understand the extended {\Lya} emission, we have
constructed a numerical model, where the quasar lies in the centre of a
large, optically thin \ion{H}{i} cloud. This model was already prosposed
in Weidinger et al.~(2004; hereafter Paper II) together with the main
conclusions. The details of our calculations were not included in that
paper, but they will now be presented here.

The quasar emission is collimated in a cone with a full opening angle
$\Psi$, and the system is seen under an inclination angle $\theta$ (see
Fig.~3a of Paper II). The \ion{H}{i} within the cone is
photoionized by the quasar UV photons, causing it to emit {\Lya}
photons when recombining. For the calculation of the extended
\Lya-emission, we follow the treatment of the optically thin case given
in Gould \& Weinberg (1996). The ionization rate of a hydrogen atom at
a distance $r$ from the quasar is
\begin{equation}
\Gamma(r) = \int_{\nu_L}^{\infty} \phi(\nu)\sigma(\nu)d\nu,
\end{equation}
where $\phi(\nu)$ is the photon flux density
\begin{equation}
\phi(\nu) = \frac{f_{\nu}(r)}{h\nu}
\end{equation}
and the \ion{H}{i} ionization cross section is
\begin{equation}
\sigma(\nu) \approx 6.3\times 10^{-18}\text{ cm}^2
\left(\frac{\nu}{\nu_L}\right)^{-2.75}.
\end{equation}
Here $h\nu_L = 13.6$ eV is the hydrogen ionization potential, and
$f_\nu(r)$ is the frequency-specific flux at a distance $r$ from the
QSO. The production rate of {\Lya} photons per unit volume is
\begin{equation}
\dot{n}_{{\rm Ly}\alpha}(r) = \eta_{\rm thin} n_{\ion{H}{i}}(r) \Gamma(r),
\end{equation}
where $\eta_{\rm thin} = 0.42$ is the fraction of recombinations that
result in a {\Lya} photon. We assume a power-law hydrogen density
profile, $n_{\ion{H}{i}}(r) = n_{{\ion{H}{i}}, 1} \left(\frac{r}{1\text{
kpc}}\right)^{-\gamma}$. By integrating along the line of sight, $l$,
we obtain the surface brightness (in erg s$^{-1}$ cm$^{-2}$
arcsec$^{-2}$)
\begin{equation}
\Sigma_{{\rm Ly}\alpha} = \frac{\int E_{{\rm Ly}\alpha}\dot{n}_{{\rm
Ly}\alpha}(r) dl}{4\pi D_L^2} \frac{D_A^2}{d\Omega}.
\end{equation}
Here $E_{{\rm Ly}\alpha} = 10.4$ eV is the energy of a {\Lya} photon,
and $D_L$ the quasar luminosity distance. $D_A$ is the angular
distance, such that $\frac{d\Omega}{D_A^2}$ is the conversion from
cm$^2$ to arcsec$^2$.

The flux emitted from the quasar close to the Lyman-limit frequency,
$\nu_L$, is given by
\begin{equation}
  f_\nu(r) =
  \begin{cases}
         f_{\nu_L,\text{ obs}}\left(\frac{D_L}{r}\right)^2
         \left(\frac{\nu}{\nu_L}\right)^{-\alpha} & \text{inside the cone,}
         \\ 0 & \text{outside the cone.}
  \end{cases}
\end{equation}
We obtain the observed Lyman-limit flux, $f_{\nu_L,\text{ obs}}$, using
the slope of the quasar spectrum and an observed flux,
$f_{\nu_0, \text{ obs}}$, on the continuum at $\nu_0$, that is
\begin{equation}
f_{\nu_L,\text{ obs}} =
f_{\nu_0, \text{ obs}}\left(\frac{\nu_L}{\nu_0}\right)^{-\alpha}.
\end{equation}

The free parameters of the model are the observed flux at $\nu_0$,
$f_{\nu_0, \text{ obs}}$, the spectral slope, $\alpha$, the redshift,
$z$, the angles $\Psi$ and $\theta$, the neutral hydrogen density at a
distance of $1$ kpc, $n_{{\ion{H}{i}}, 1}$, and the density slope,
$\gamma$.

\subsection{Model parameters} 

Several free parameters are determined directly via observations.
We measure the observed continuum flux at a given point in
the spectrum. For the spectral index $\alpha$ we use the value found in
the flux calibration in Sect.~\ref{sec:cal}. The obtained values for
these parameters are listed in Table~\ref{tab:qsim}. In Paper II a
method to obtain a relation between the opening angle and the
inclination angle is described. The free parameters in our model are
now reduced to the opening angle, $\Psi$, the hydrogen density scale,
$n_{{\ion{H}{i}}, 1}$, and the density slope, $\gamma$.

The numerical implementation of the model was carried out by
incorporating the cone into a cubic grid inclined at an angle
$\theta$. The {\Lya} production rate was calculated in each grid point
within the cone, and the process of integrating along the line of sight
was simply to sum along the grid $z$-axis.

\begin{table}[htb]
\caption{Input parameters of the model.}
\label{tab:qsim}
\centering\begin{tabular}{l c c}
\hline\hline
 Description      & Symbol   & Value      \\\hline
 Observed flux    & $f_{\lambda_0, \text{ obs}}$    & {\scriptsize $1.3\times
 10^{-16}$           erg s$^{-1}$ cm$^{-2}$ \AA$^{-1}$} \\
 Reference wavelength & $\lambda_0$ & $5946$ \AA \\
 Spectral slope   & $\alpha$ & $0.65$     \\
 Redshift         & $z$      & $3.041$    \\\hline
\end{tabular}
\end{table}

\begin{figure}[tb]
\centering\epsfig{file=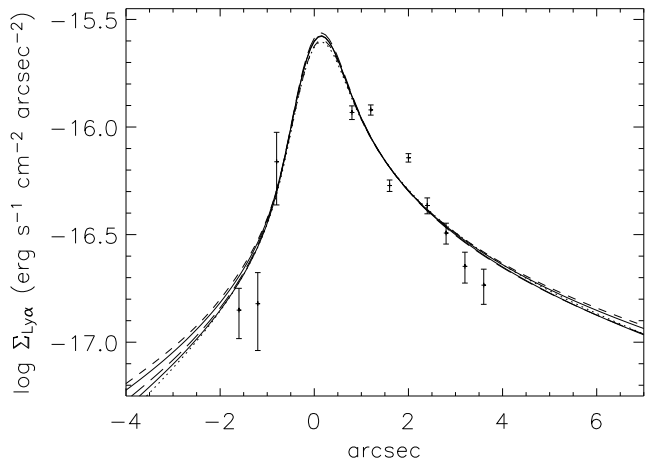,width=8.8cm}
\caption{Best-fit surface brightness profiles as seen through a $1.2$
arcsec wide slit with a seeing of $0.7$ arcsec for opening angles of
$90^\circ$ (dotted), $110^\circ$ (solid), $130^\circ$ (long dashes),
$150^\circ$ (solid), and $170^\circ$ (short dashes). The model
parameters are listed in Table~\ref{tab:qsimbestfit}. Overplotted is
the measured surface brightness profile of S6 shown in
Fig.~\ref{fig:result}b. Note that three marginal measurements at
negative offsets are included. }
\label{fig:SBprofile}
\end{figure}

\begin{table}[Hhtb]
\caption{Best-fit values for $\theta$, $n_{{\ion{H}{i}}, 1}$, $\gamma$, and
  $M_{200}$ for various opening angles.}
\label{tab:qsimbestfit}
\centering\begin{tabular}{c c c c c}
\hline\hline
$\Psi$ & $\theta$ & $n_{{\ion{H}{i}}, 1}$ & $\gamma$ & $M_{200}$ \\
 &       & \scriptsize$(10^{-8}\text{ cm}^{-3})$ & &
 \scriptsize$(10^{12} M_\odot)$ \\\hline
$90^\circ$  & $22.5^\circ$ & $9.8$ & $0.018$ & 2.6\\
$110^\circ$ & $27.2^\circ$ & $9.3$ & $0.072$ & 3.5\\
$130^\circ$ & $31.8^\circ$ & $8.2$ & $0.093$ & 5.3\\
$150^\circ$ & $36.5^\circ$ & $6.4$ & $0.064$ & 6.5\\
$170^\circ$ & $41.1^\circ$ & $5.5$ & $0.054$ & 6.5\\\hline
\end{tabular}
\end{table}

\subsection{Results of the model}

Using the parameters listed in Table~\ref{tab:qsim}, and assuming a
given opening angle and corresponding best-fit inclination angle, we
now fit the model to the observed surface brightness profile, leaving
the hydrogen density scale, $n_{{\ion{H}{i}}, 1}$, and slope, $\gamma$,
as free parameters. The best-fit values for each opening angle are
listed in Table~\ref{tab:qsimbestfit}, and the corresponding surface
brightness profiles are plotted in Fig.~\ref{fig:SBprofile}.

The numerical model described in Paper II and above is purely
geometrical. Assuming in addition that the gas is in free fall into a
dark matter (DM) halo introduces a velocity field in the gas. In Paper
II this was used to fit the virial mass, $M_{200}$, to the observed
velocity profile. The obtained virial masses are also listed in
Table~\ref{tab:qsimbestfit}. The conclusions are presented in Paper II.

\section{Discussion}\label{sec:q1205:5} Returning to the main topic of
the origin of the extended {\Lya} emission, there are several
elements of this puzzle that now can to be pieced
together. There is a LLS at $z=3.0336 \pm 0.0004$ (absorber K, see
Table \ref{tab:systems}), an additional absorber at $z=3.039 \pm 0.002$
(K1, see Sect.~\ref{sec:abssystems}), the systemic redshift of the
quasar is $z=3.041 \pm 0.001$ (Sect.~\ref{sec:systemicz}), the absorber
L at $z=3.0485 \pm 0.0005$ (Table~\ref{tab:systems}) and finally the
extended {\Lya} emission S6 at $z\approx 3.049$ (see caption of
Fig.~\ref{fig:result}c-d). We assume that the high-ionization,
line locked systems I and J are intrinsic to the central engine
and will not consider them here.

\subsection{The origin of the extended {\Lya} emission}
The separation of the low- and high-ionization lines in the LLS
suggests that it is heated by the quasar UV photons and
expanding. However, the large velocity offset between the LLS and the
extended {\Lya} emission ($\sim 1000$ km s$^{-1}$) makes it very
unlikely that the two are physically associated as previously thought
(Paper I). Since the LLS absorbs all UV photons, the extended emission
must be located between the LLS and the QSO, which makes an association
between the extended {\Lya} emission and the QSO the most likely
one. It was suggested by Haiman \& Rees (2001) that neutral gas
falling into the dark matter halo around a quasar could be photoionized
by the quasar UV flux, causing the gas to emit {\Lya} photons.

Interpreting the observations in the Haiman \& Rees (2001) picture as
described in the numerical model (Sect.~\ref{sec:model}; Paper II), we
identify K1 as absorption due to a large hydrogen cloud around the
quasar, in which case the redshift of K1 should be close to that
of the quasar. A part of this large cloud is pulled into the DM
halo of the QSO where it is photoionized and gives rise to the {\Lya}
emission of S6. S6 therefore lies between us and the quasar, and
the higher redshift is due to infall. The velocity of S6 is $\sim 500$
km s$^{-1}$ relative to the quasar and the surrounding cloud (K1). The
absorption system L is located somewhere between the emitting and the
absorbing part of the hydrogen cloud. There are at least two
possibilities for placing the LLS. \emph{i)} It is very close to the
quasar and moving at high velocity ($\sim -500$ km s$^{-1}$). In this
case the LLS has to be very small in order for the majority of the UV
photons to pass by and photoionize the surrounding hydrogen
cloud. \emph{ii)} The LLS is sufficiently distant for the {\Lya}
photons produced in the cloud surrounding the quasar to be redshifted
out of the resonance wavelength and pass through the LLS
unhindered. Because the low- and high-ionization lines in the LLS are
only mildly separated in velocity space ($\sim 60$ km s$^{-1}$)
we favour the latter possibility.

\begin{table*}[htb]
\caption{Redshifts, fluxes and luminosities of the {\Lya} emission from
the host galaxies of radio-quiet quasars. }\label{tab:lum}
\centering\begin{tabular}{l c c c >{\scriptsize}l}
\hline\hline
Quasar & $z$ & {\Lya} flux &{\Lya} luminosity & {\normalsize Reference}\\ & & {\scriptsize
$(10^{-17}$~erg~s$^{-1}$~cm$^{-2})$} & {\scriptsize
$(10^{42}$~erg~s$^{-1})$} & \\\hline 
 \object{Q 0054-284}   & $3.616$ & $ 6$  & $7.3$   & Bremer et al.~(1992)  \\
 \object{Q 0055-26}    & $3.656$ & $ 8$  & $ 10$   & Bremer et al.~(1992)  \\
 \object{Q 1548+0917}  & $2.746$ & $ 47$ & $ 29$   & Steidel et al.~(1991) \\
 \object{Q 1205-30}    & $3.041$ & $ 70$ & $ 56$   & This work             \\
 \object{BR 1202-0725} & $4.69$  & $ 26$ & $ 58$   & Hu et al.~(1996), Petitjean et al.~(1996)\\
\hline
\end{tabular}
\end{table*}

The extended {\Lya} emission could possibly be explained by other
scenarios rather than the projected ionization cone. However, the
combination of imaging and spectroscopic data enables us to rule out
most of these other scenarios (Paper II). \emph{i)} Jets
are believed to be present in radio-quiet quasars. They are predicted
to extend out to only $\sim 0.1$ kpc (Blundell et al.~2003), more than
two orders of magnitude less than the $\sim 30$ kpc extent of the
emission around \object{Q1205-30}. \emph{ii)} Outflowing galactic winds are
generally thought to be triggered by the cumulative effect of many
supernovae exploding inside the galaxy, which would metal-enrich the
outflowing gas. Around radio-loud quasars this is typically seen as
extended \ion{C}{iv} emission with strength $7-10\%$ of that of the
extended {\Lya} line (Heckman et al.~1991b). Our detection limit at the
position of redshifted \ion{C}{iv} is $4\times 10^{-18}$ erg s$^{-1}$
cm$^{-2}$ arcsec$^{-2}$ ($3\sigma$), which would have enabled us to
detect the typical \ion{C}{iv} line seen around some radio-loud quasars
(a strength of $10\%$ of the {\Lya} line corresponds to $7\times
10^{-18}$ erg s$^{-1}$ cm$^{-2}$ arcsec$^{-2}$). The fact that we do
not detect this line makes it unlikely that it is a supernova powered
outflow. The most plausible explanation that remains is cosmological
infall of hydrogen.

\subsection{Extending the sample}

To date only a handful of detections of extended {\Lya} emission around
RQQs has been reported (Steidel et al.~1991; Bremer et al.~1992; Hu et
al.~1996; Petitjean et al.~1996; Bunker et al.~2003). Hu et al.~(1991)
find no extended {\Lya} emission in their sample of seven radio-quiet
quasars down to a limiting flux of $2\times
10^{-16}$~erg~s$^{-1}$~cm$^{-2}$. We have compiled a list of fluxes and
luminosities for the detections of extended {\Lya} emission (see Table
\ref{tab:lum}). We measure an average {\Lya} surface brightness of
$\sim 7\times 10^{-17}$ erg s$^{-1}$ cm$^{-2}$ arcsec$^{-2}$ around
\object{Q1205-30}. Assuming a spatial extent of $10$ arcsec$^2$, we infer a
{\Lya} flux of $\sim 7\times 10^{-16}$ erg s$^{-1}$ cm$^{-2}$. The
luminosity in the {\Lya} line of S6 is at the high end of the ones
listed in Table \ref{tab:lum}.

It is important for the study of the link between galaxy and quasar
formation to understand how frequent and under which circumstances
extended {\Lya} emission arises around RQQs. The most efficient method
to detect {\Lya} emission around QSOs is narrow band imaging. The study
of extended {\Lya} emission can successfully be combined with narrow
band searches for {\Lya} emitting (proto)-galaxies around QSOs
(e.g.~Fynbo et al.~2001; Fynbo et al.~2003). The follow-up multi-object
spectroscopy needed to confirm candidate {\Lya} emitters may
conveniently be utilized to study any extended {\Lya} emission
associated with the QSO. Alternatively, integral field spectroscopy is
a promising method for very detailed studies of extended emission at
high redshifts (Bower et al.~2004). The method makes it possible to map
out the entire velocity field of the extended emission, strongly
constraining any model.

It is imperative for any search for extended {\Lya} emission around
RQQs to go to very deep detection limits. Haiman {\&} Rees (2001)
predict that haloes of infalling gas around quasars should be seen in
{\Lya} emission with a typical surface brightness around $10^{-18} -
10^{-17} $~erg~s$^{-1} $~cm$^{-2} $~arcsec$^{-2}$ and typical angular
sizes between $2-3$~arcsec, i.e.~with typical fluxes around $3\times
10^{-18} - 7\times 10^{-17} $~erg~s$^{-1} $~cm$^{-2}$. This  limit
has only been reached for very few surveys.

A larger sample of QSOs with extended {\Lya} emission will make it
possible to address primary issues like morphology, environment,
luminosity function etc. The corresponding DM halo masses obtained in a
similar fashion as in Paper II may be compared to black hole masses
obtained via the correlation between $M_{\rm BH}$ and quasar emission
line widths (Vestergaard 2002; McLure \& Jarvis 2002). A $M_{\rm DM} -
M_{\rm BH}$ correlation could provide a powerful consistency check of
$N$-body hydrodynamical simulations.

We are currently looking for extended {\Lya} emission in a small sample
of quasars using an analysis similar to what has been employed in this
paper.

\begin{acknowledgements}
MW acknowledges support from ESO's Director General's Discretionary
Fund. We wish to thank Pall Jakobsson for helpful discussions of
gravitational lensing, and Henning J{\o}rgensen for useful comments on
our manuscript. It is a great pleasure to thank the referee Cedric
Ledoux for his large effort and very helpful comments.
\end{acknowledgements}

\end{document}